\documentclass[namedreferences]{kluwer}    

\usepackage[dvips]{graphicx}
\usepackage{makeidx}

\makeindex

\begin{document}                                                                                   
\begin{article}
\begin{opening}         
\title{An Overview of The VERITAS Prototype Telescope And Camera} 
\author{Peter \surname{Cogan for the VERITAS \footnote{The VERITAS collaboration consists of universities and institutions from Ireland, UK, US \& Canada. See http://veritas.sao.arizona.edu/VERITAS\_members.html for a full listing.}
  Collaboration}}  
\runningauthor{Peter Cogan}
\runningtitle{An Overview of The VERITAS Prototype Telescope And Camera}
\institute{University College Dublin, National University of Ireland}

\date{May 31, 2004}

\begin{abstract}
VERITAS (the Very Energetic Radiation Imaging Telescope Array System)
is the next generation ground-based gamma-ray observatory that is
being built in southern Arizona by a collaboration of ten institutions
in Canada, Ireland, the U.K. and the U.S.A. VERITAS is designed to
operate in the range from 50 GeV to 50 TeV with optimal sensitivity
near 200 GeV; it will effectively overlap with the next generation of
space-based gamma-ray telescopes. The first phase of
VERITAS, consisting of four telescopes of 12 m aperture, will be
operational by the time of the GLAST launch in 2007. Eventually the
array will be expanded to include the full array of seven telescopes
on a filled hexagonal grid of side 80 m. A prototype VERITAS telescope
with a reduced number of mirrors and signal channels has been
built. Its design and performance is described here.  The prototype is
scheduled to be upgraded to a full 499 pixel camera with 350 mirrors
during the autumn of 2004.
\end{abstract}

\end{opening}

\section{Introduction - Very High Energy $\gamma$-ray Astronomy}
Very High Energy (VHE) $\gamma$-ray astronomy is the study of photons
in the 50 GeV to 50 TeV energy range. Such photons are only produced by the
most exotic objects in the universe with Active Galactic Nuclei (AGN) and
Super-Nova Remnants (SNR) among the source categories detected
thus far. Given that VHE $\gamma$-rays are at the extreme end of the
electromagnetic spectra of these objects, they provide excellent
constraints on their emission models. The study of VHE $\gamma$-rays
may also be used to help determine the source of cosmic rays, measure the
density of the extragalactic background infrared (IR) radiation and measure the
magnetic fields in the shells and nebulae of SNRs. The effective
detection of VHE photons is impractical with space-based telescopes
due to the low flux of photons and the small collection area of
such telescopes. Instead, VHE $\gamma$-rays may be indirectly
quantified by measuring the Extensive Air Shower (EAS) produced
when a $\gamma$-ray strikes the earth's atmosphere.

\section{The Imaging Atmospheric Cherenkov Technique}
Upon interacting with the earth's atmosphere VHE $\gamma$-rays produce
cascades of charged particles known as extensive air showers
(EAS). These may be detected indirectly at ground level by instruments
capable of recording the Cherenkov radiation emitted by the
relativistic shower constituents. Unfortunately there exists an almost
overwhelming background of EAS produced by charged cosmic rays and an
extremely efficient background rejection method is required. The
Imaging Atmospheric Cherenkov Technique (IACT)
\cite{weekesandturver} allows discrimination between cosmic-ray and
gamma-ray initiated EAS by recording the images of the Cherenkov
light emitted by the EAS as they develop in the
atmosphere \cite{hillas}. Off-line image analysis techniques are
employed to exploit the subtle differences in the physics of the
hadronic and electromagnetic EAS, resulting in the rejection of over
99.7\% of the background. The imaging cameras are made from arrays of
Photo-Multiplier Tubes (PMTs) located in the focal planes of large
optical reflectors.

\section{VERITAS}
The Whipple 10m telescope was the first IACT telescope to be operated
\cite{cawley}. It detected its first VHE $\gamma$-ray source, the Crab
Nebula, in 1989 \cite{weekesetal}. Since that time several other
collaborations have built IACT telescopes and the catalogue of
detected $\gamma$-ray sources has blossomed to nearly 20 objects
\cite{horan}. The angular and energy resolution of the Imaging
Atmospheric Cherenkov Technique can be improved by utilising an
array of such telescopes. This approach will also lower the energy
threshold, thus narrowing the energy gap between space-based and
ground-based telescopes. The operation of multiple IACT telescopes in
stereo mode has been demonstrated by the HEGRA group \cite{konopelko}.

\begin{figure}
\centerline{\includegraphics[width=15pc]{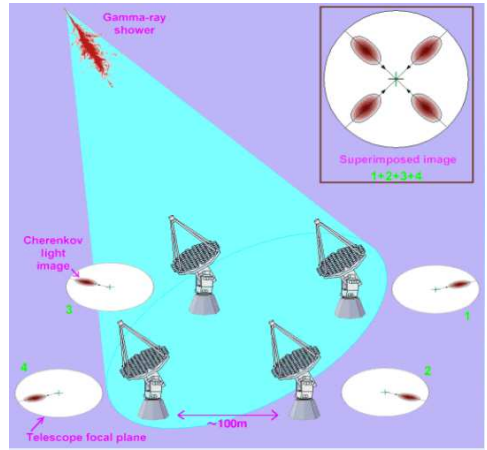}}
\caption{Several telescopes may lie in the light pool generated by an EAS. This is the key to stereoscopic imaging.(Holder, U.Leeds)}
\end{figure}

 The Very Energetic Radiation Imaging Telescope Array System (VERITAS)
\cite{weekesetal2} is a next generation IACT telescope array currently
being built in Arizona by the VERITAS collaboration. The first stage
in building VERITAS was to construct a single prototype telescope
\cite{wakelyetal}. This telescope was erected during the summer of 2003
at the basecamp of the Fred Lawerence Whipple Observatory in
Arizona. The primary mission of the prototype telescope was to field
test the new technologies being employed by VERITAS and to test system
integration. The prototype telescope operated from September 2003
until April 2004.

\section{The VERITAS Prototype}
\subsection{Telescope and Optics}
The VERITAS prototype telescope was built using a custom-designed,
welded steel, space-frame Optical Support Structure (OSS) mounted on a
commercial positioner \cite{gibbsetal}. The telescope uses the Davies
Cotton \cite{daviescotton} reflector design. This design comprises
multiple identical small mirrors arranged such that they mimic a large
single reflector. The most significant advantage of this design is
that it removes the requirement to build a single large mirror - which
would be problematic in terms of cost, weight and gravitational
slumping. The Davies Cotton design is cheap, the mirrors light and
individually adjustable. If a single mirror is damaged it can be
easily replaced. The most significant disadvantage of the Davies
Cotton design is that it is not isochronous. This means that a time
spread is introduced to the light pulse arrival time at the
camera. The VERITAS telescope's focal length is 12m, making it an f/1
system for a 12m aperture. This is a significant improvement over the
Whipple telescope which is f/0.7 as it will reduce the effect of
optical aberrations and is required in order to match the angular size
of the PMTs (0.15$^{\circ}$). It also reduces the light pulse arrival time
spread to 4ns.

The total mass of the OSS, mirrors, counterweights and camera is
estimated to be 16,000 kg. RPM has provided a positioner,
model PG-4003, to meet the requirements of the VERITAS telescope. The
positioner has a design azimuth slew speed of 1$^{\circ}$/s and a design
elevation slew speed of 0.5$^{\circ}$/s, although it was being exercised at
more conservative speeds while its performance was evaluated.

\vspace{-0em}

\begin{figure}[H]
\tabcapfont
\centerline{%
\begin{tabular}{c@{\hspace{3pc}}c}
\includegraphics[width=15pc]{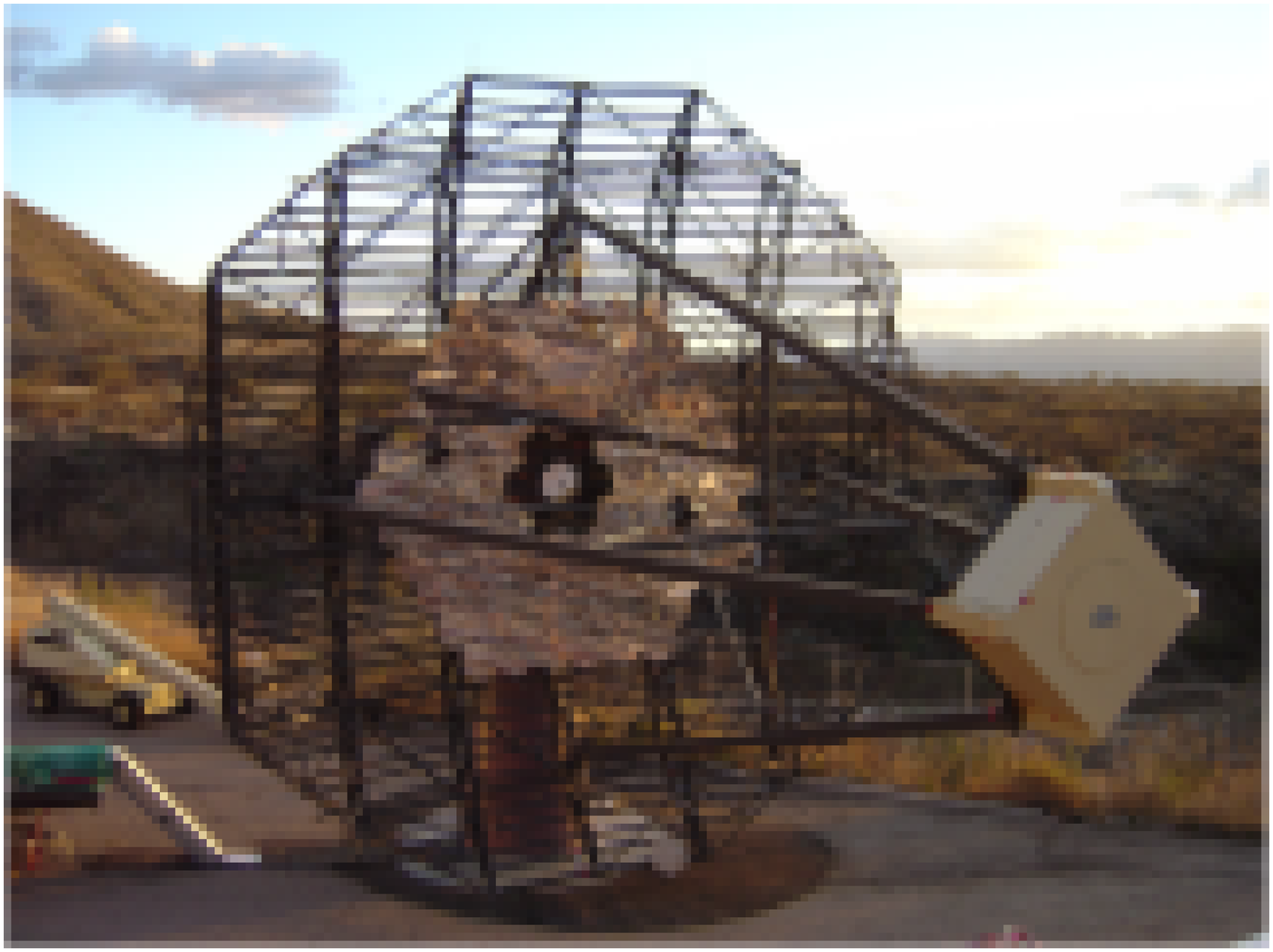} &
\includegraphics[width=15pc]{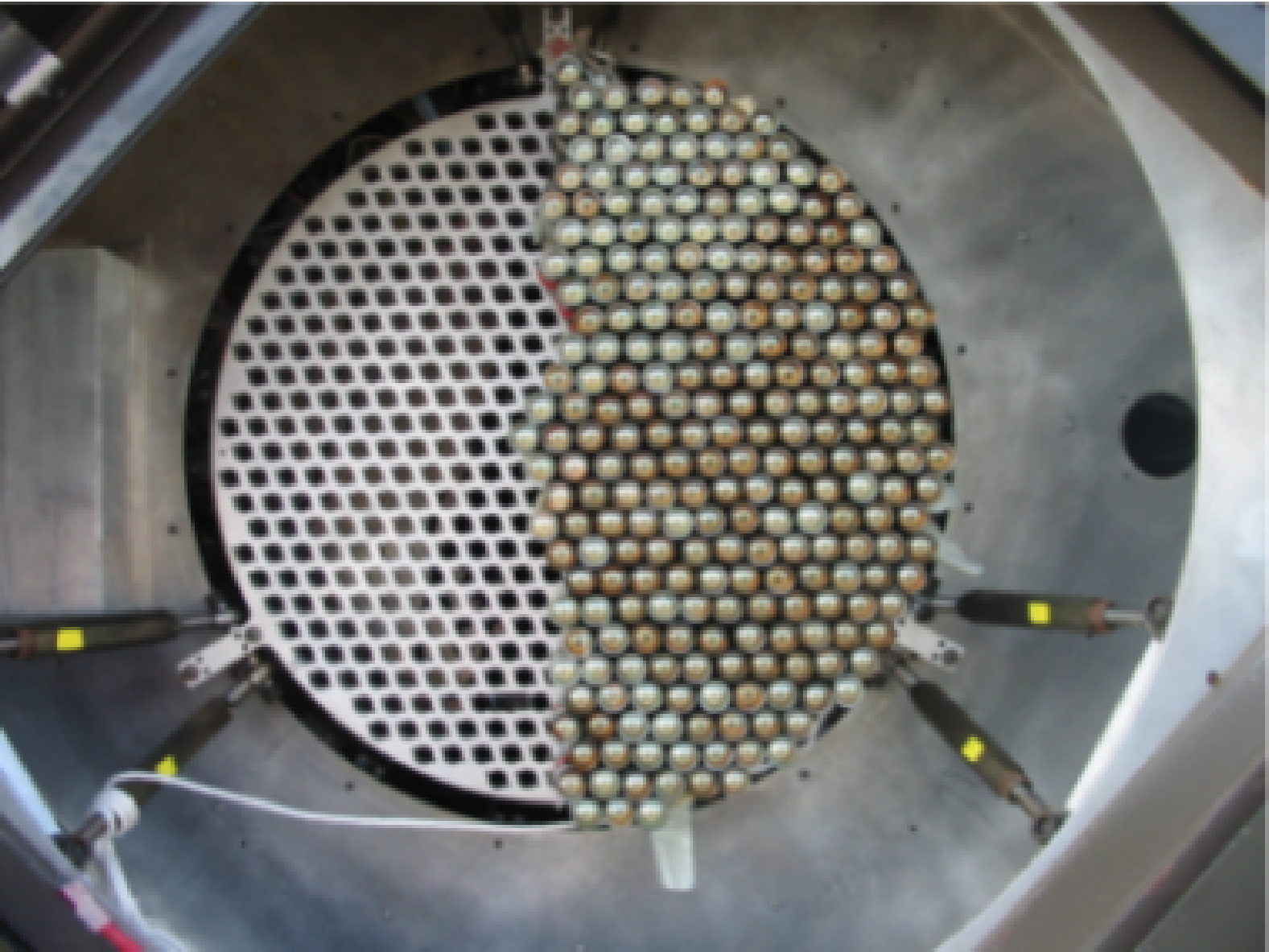} \\
~~ ~~
\end{tabular}}
\caption{Left:The VERTIAS Prototype. Right: The Prototype Camera}
\end{figure}



Although the prototype employed a full sized OSS, it only held 84
mirrors giving a surface area of 34 m$^2$. Each full VERITAS telescope
will hold 350 mirrors giving a total mirror area of 140 m$^2$. The
optical qualities of the VERITAS prototype mirrors were tested at the
Whipple Observatory basecamp. They were found to exceed design
specifications in terms of reflectivity and curvature. Mirror
alignment is achieved using a laser based Mirror Alignment System. The
Point Spread Function of the prototype telescope, which is a measure
of the mirror alignment, was 0.05$^{\circ}$ which is within design
specifications.


\subsection{Camera}
In order to detect the extremely faint and brief Cherenkov flash, a
low noise, high gain photon counting device with a fast (2.5ns)
risetime is required. Despite steady advances in the field of solid
state detectors, only Photo Multiplier Tubes (PMTs) currently meet
these requirements. A VERITAS camera will comprise 499 1 1/8 inch
Photonis XP 2970 PMTs. The PMTs are supplemented with a high bandwidth
preamplifier integrated into the PMT base. A charge-injection system
is also present in the camera for calibration and diagnostic
purposes. This system injects pulsed charges into the preamplifier
which simulates Cherenkov events. This can be used to test the
electronics and data acquisition systems downstream of the PMTs in a
controlled manner without activating the PMTs. Current monitoring
systems in the camera provide a measurement of the brightness of the
field of view in each pixel and may be used to check tracking. Each
full VERITAS camera will also contain environmental sensors. The High
Voltage (HV) for the PMTs is provided by a multichannel modular
commercial power supply (CAEN SY1527/A1932). The VERITAS prototype
camera consisted of 255 Hamamatsu PMTs in a `half-moon'
configuration. Despite the reduced camera size, the VERITAS prototype
performed excellently as a full IACT telescope.

\section{Trigger and Calibration Systems}

VERITAS utilises a flexible three-level trigger system which allows
full-array, sub-array and individual telescope operation. Trigger
level one consists of individually programmable constant-fraction
discriminators (CFDs) which determine whether an individual pixel has
triggered.  The level two trigger is a topological hardware trigger
which can discriminate between compact Cherenkov events and random
night-sky or afterpulse-induced events \cite{bradburyetal}. Trigger levels one and
two constitute the telescope-level trigger while the level three
system receives inputs from all the telescopes and generates
array-level triggers, and event numbers, based on a geometry-adjusted
multiplicity condition.  The VERITAS prototype has operated with full
implementations of the level one and level two triggers for 250
channels but with a simplified level 3 trigger as sophisticated
multi-telescope triggering is not yet required.

Calibration systems installed on the VERITAS prototype telescope
include a charge injection system and a nitrogen/dye laser flasher. In
addition, atmospheric monitoring stations, a mid-infra-red pyrometer
system, and CCD sky quality monitoring cameras are under development.

\section{Signal Digitisation and Data Acquisition}
The signals from each of the PMTs are digitised by a custom-designed
500 MSPS flash-analogue-to-digital converter (FADC) system
\cite{buckleyetal}. The prototype telescope has 25 FADC boards, containing
10 channels each, situated in two VME crates.  A third VME crate
contains GPS clocks and ancillary modules. Each FADC channel has a
dynamic range of $\sim$11 bits and a memory depth of $\sim$8~$\mu$s.
The FADCs have zero-suppression capabilities so that channels whose
signal do not exceed a programmable threshold do not appear in the
data stream.

\begin{figure}
\centerline{\includegraphics[width=20pc]{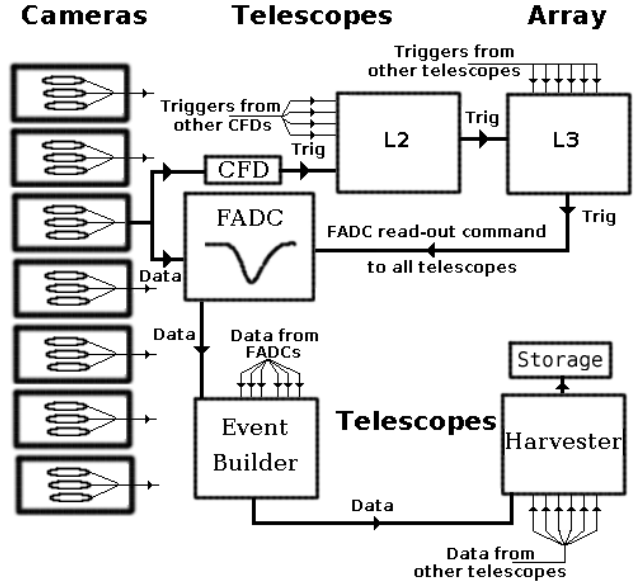}}
\caption{A rough map of the data/trigger chain
for the VERITAS array. For the prototype there was only one
camera/telescope however the array trigger section was present.}
\end{figure}
 
With the FADCs, VERITAS can sample the charge deposited in each pixel
every 2ns. This provides a pulse profile for each pixel for each
event. As well as improving the signal to noise ratio, eliminating the
need for delay cables and reducing deadtime, the pulse profile may
provide a new technique for discriminating against hadronic events
using the time evolution of the EAS.

Readout of the FADCs is accomplished using VMIC Pentium crate
computers running the Linux operating system. Each crate computer
buffers multiple event fragments until an optimal depth has been
reached, when they are then transferred via Scalable Coherent
Interface to the telescope event-building system. The telescope
event-building system consists of a dual-processor Xeon server running
Linux and the data acquisition software is implemented in C++. Built
telescope events are sent via ethernet to the harvester machine, where
they are combined into array events and analysed, and stored locally
on disk for redundancy. The DACQ software is controlled from the
central Array-Control system via a CORBA interface.

This integrated data acquisition system has been thoroughly tested on
the prototype telescope and found to perform reliably. During the
prototype telescope evaluation stage 64 samples were read out for
every channel with each event, resulting in a data size of $\sim$16.5
kBytes per event, at a rate of $\sim$30~Hz. The full VERITAS
telescopes will operate with a considerably reduced event size due to
zero-suppression and a reduced number of FADC samples per event.


\section{Early Results}
The VERITAS prototype operated from September 2003 until the end of
April 2004. While much of this time was spent testing systems and
system integration, a significant amount of quality data was taken on
the well established TeV source Markarian 421 (Mrk421). During this
period, the nearby Whipple 10m telescope observed this BL Lac in a
flaring state, allowing a meaningful comparison in gamma-ray rate
between the Whipple 10m telescope and the VERITAS prototype. The
prototype detected Mrk421 at 20.5$\sigma$ over 19.2 hours of
observations. During the same period, the prototype telescope's average rate was
0.8 $\gamma$ /min and the Whipple 10m observatory's was 7
$\gamma$ /min.





\begin{figure}[H]
\tabcapfont
\centerline{%
\begin{tabular}{c@{\hspace{3pc}}c}
\includegraphics[width=17pc, height=13pc]{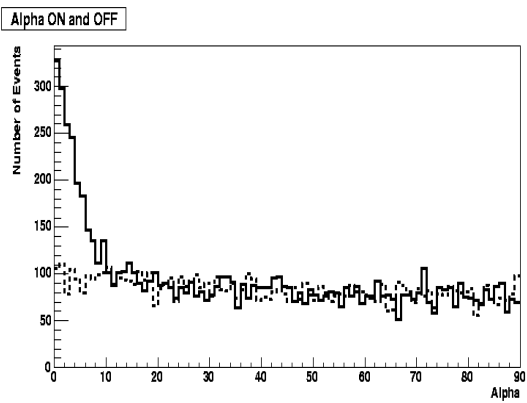} &
\includegraphics[width=17pc]{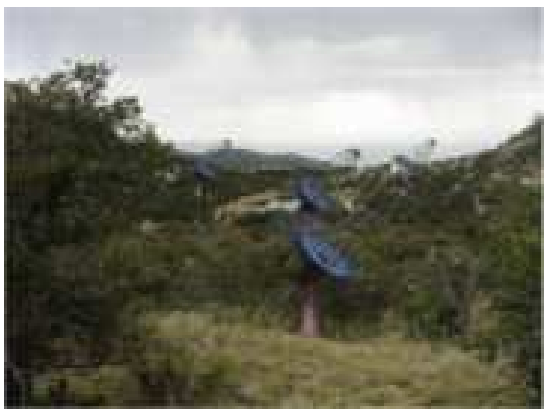} \\
~~ ~~
\end{tabular}}
\caption{Left: An alpha plot indicating the detection of Mrk421 (Holder, U. Leeds). Right: An impression of the complete VERITAS-4 array at Horseshoe Canyon, Arizona.}
\end{figure}

\section{Schedule}
During the summer of 2004, the prototype camera will be completely
disassembled and the new Photonis pixels installed . The remainder of
the 350 mirrors will be mounted. When these upgrades are completed the
prototype will be redesignated Telescope 1 and will operate at the
basecamp of the Whipple Observatory for two years. In the interim, the
VERITAS site at Horseshoe Canyon will be prepared and three telescopes
identical to Telescope 1 will be constructed. Upon their completion,
Telescope 1 will be moved to Horseshoe Canyon, completing the
VERITAS-4 array. This is expected to occur before the summer of 2006.




\end{article}
\end{document}